\documentclass[12pt]{article} 
\usepackage{amsmath}
\usepackage{psfrag,epsf} 
\usepackage{natbib}
\usepackage{float}   
\usepackage[dvips]{graphicx} 
\usepackage{booktabs} 
\usepackage[utf8x]{inputenc}
\usepackage[english]{babel}
\usepackage{natbib}
\usepackage[fixamsmath]{mathtools}
\usepackage{mathtools}  
\usepackage{times}
\usepackage{lipsum}
\usepackage{float}
\usepackage{enumitem}
\usepackage{caption}
\usepackage{amsfonts}
\usepackage{forest}
\usepackage{siunitx}
\usepackage{adjustbox}
\usepackage{tikz} 
\usepackage[labelfont=bf]{caption}  
\usepackage{hyperref}
\usepackage{url} 

\newcommand{\blind}{0}

\begin{document}

\def\spacingset#1{\renewcommand{\baselinestretch}%
{#1}\small\normalsize} \spacingset{1}


\if0\blind
{
  \title{\bf Additivity Assessment in Nonparametric Models Using Ratio of Pseudo Marginal Likelihoods}
  \author{\bf{Bonifride Tuyishimire$^1$, Brent R. Logan$^1$ and Purushottam W. Laud$^1$}
 		\hspace{.2cm}\\  
    $^1$Division of Biostatistics, Medical College of Wisconsin}
  \maketitle
} \fi

\if1\blind
{
  \bigskip
  \bigskip
  \bigskip
  \begin{center}
    {\LARGE\bf Title}
\end{center}
  \medskip
} \fi

\bigskip
 
\begin{abstract}
Nonparametric regression models such as Bayesian Additive Regression Trees (BART) can be useful in fitting flexible functions of a set of covariates to a response, while accounting for nonlinearities and interactions. However, they are often cumbersome to interpret. Breaking down the function into additive components, if appropriate, could simplify the interpretation and improve the utility of the model. On the other hand, establishing nonadditivity can be useful in determining the need for individualized predictions and treatment selection. Testing additivity of single covariates in nonparametric regression models has been extensively studied. However, additivity assessment of nonparametric functions of disjoint sets of variables has not received as much attention. We propose a method for detection of nonadditivity of two disjoint sets of variables by fitting the sum of two BART models, each using its own set of variables. We then compare the pseudo marginal likelihood (PsML) of this sum-of- BARTs model vs. a single-BART model with all the variables together, in a ratio known as Pseudo Bayes Factor (PsBF) to allow for model assessment between the additive and nonadditive models. A special case of our method checks additivity between one variable of interest and another set of variables, where the additive model allows for direct interpretation of the variable of interest while adjusting for the remaining variables in a flexible, nonparametric manner. We extended the above approaches to allow a binary response using a logit link. We also propose a systematic way to design simulations that are used in additivity assessment. In simulation studies, PsBF showed better performance compared to out-of-sample prediction error in terms of the probability of detecting the correct model, while avoiding computationally expensive cross-validation and providing an interpretable criterion for model selection. We applied our approach to two different examples; one with a continuous and the other with a binary outcome. In the first example, we use data from a Type 2 diabetes study to investigate additivity between health and demographic variables, and between a treatment and all other variables. In the second example, we use data from patients undergoing a hematopoietic cell transplant and assess whether the effect of patient, donor and disease factors on their one year survival is additive with the treatment effect. 
\end{abstract}

\noindent%
\vfill

\newpage
\spacingset{1.45} 
\section{Introduction}
\label{sec:intro} 

Bayesian Additive Regression Trees (BART) \citep{bart2010} is a nonparametric, Bayesian model that uses sum-of-trees to flexibly fit relationships between an outcome and a set of covariates. BART automatically accounts for nonlinear relationships and incorporates interactions. The excellent predictive performance of BART and easy accessibility of BART software \citep{bart2010,bartmanual,Line18,SparLoga16, LogaSpar17,KapelBleich16,ChipMacC16} have lead to its popular use. However, the resulting regression function can be very cumbersome, making the task of extracting and interpreting individual covariate effects more complicated. 

For example, in this article we consider data from Type 2 diabetes patients obtained from the Clinical Research Data Warehouse of the Froedtert Health and the Medical College of Wisconsin. Hemoglobin A1C (HbA1C) measures the severity of Diabetes, and lowering it to a healthier level, usually 7\%, is the primary goal of physicians and patients \citep{KoenCera80,Ada14}. HbA1C measurements, 6 months apart, were taken on 2673 patients. Many factors contribute to the lowering of HbA1C; these include health variables such as baseline HbA1C, depression, hypertension, BMI, HDL, LDL, systolic blood pressure (BP), diastolic BP, and demographic variables such as age, gender, race, insurance, marital status, and religion. Treatments such as metformin and insulin are commonly used in controlling HbA1C levels in type 2 diabetes patients. The treatment variable used in this article is an indicator of whether a patient was taking a treatment for diabetes or not regardless of which treatment they took. In addition to being able to predict HbA1C accurately given these variables, researchers are also interested in understanding the effect of different variables on HbA1C. As fitting a BART model with all of the above variables would result in a complicated function, breaking it down into additive parts, if appropriate, could simplify the interpretation and improve the practical utility of the model. This amounts to fitting a model with additive nonparametric functions with each function fit with its own set of covariates. A special case of this additive model fits a parametric model to one of the additive components with one (treatment for example) or more variables, while allowing the remaining variables to have a flexible form. If additive, this results in a more interpretable treatment effect; if not additive, this indicates a need for individualized treatment considerations.\\

 Some approaches for assessing additivity of two or more univariate predictors have been proposed \citep{zhang2016, eubank1995, barry1993}. However, formal methods of additivity assessment of nonparametric functions of disjoint sets of multiple variables  are still lacking; such a method would be needed, for example, to assess the additivity of health and demographic factors in the type 2 diabetes example. Out-of-sample prediction error (OSPE) has been used as a criterion to compare the prediction performance of one model versus another. However, it is computationally intensive to calculate. Log Pseudo Marginal Likelihood (LPML) \citep{GeisEddy79,BranJohn15,ChrisJohn11} is one of the criteria used in the Bayesian framework to compare two models with vague, or even improper, priors based on their prediction performance. LPML, based on the predictive distribution, is easy to compute and has been also widely adopted for model selection \citep{AbboMaro17, AnsaJedi00, BarcDeIo17, BootEile11, CardTemp04, ChanGian06, ChenHans14, HansTim2007, HeydFu16, HuHuang2015}. We propose using the LPML criterion to test nonadditivity in BART models. \\

This paper is outlined as follows: In section \ref{sec:review}, we review the BART models for continuous and binary outcomes. We also review the Log Pseudo Marginal Likelihood (LPML) criterion and Out-Of-Sample Prediction error (OSPE) criteria and how they are used for model comparison. Section \ref{sec:add} contains the proposed additive BARTs models and additive treatment BART models for both binary and continuous outcomes. This section also describes how LPML and OSPE are used to assess additivity in BARTs models. In section \ref{sim:design}, we present a systematic approach for designing simulations used in nonadditivity testing; this is based on formulating the strength of the interaction term relative to the other components in the model. Section \ref{sims} contains our simulation settings and results. In section \ref{sec:application}, we apply our approach to data from a Type 2 diabetes study to investigate additivity between health and demographic variables, and between a treatment and all other variables. We also apply the proposed approach to data from patients undergoing a hematopoietic cell transplant and assess whether the effect of their characteristics on their one year survival is additive with the treatment effect. We end the paper with a conclusion and a discussion of possible extensions of this approach.

\section{Background}
\label{sec:review}

Here we review BART models for continuous and binary outcomes as well as two model prediction performance criteria used in this article.

\subsection{Bayesian Additive Regression Trees (BART); continuous response}
\label{sec:bart}

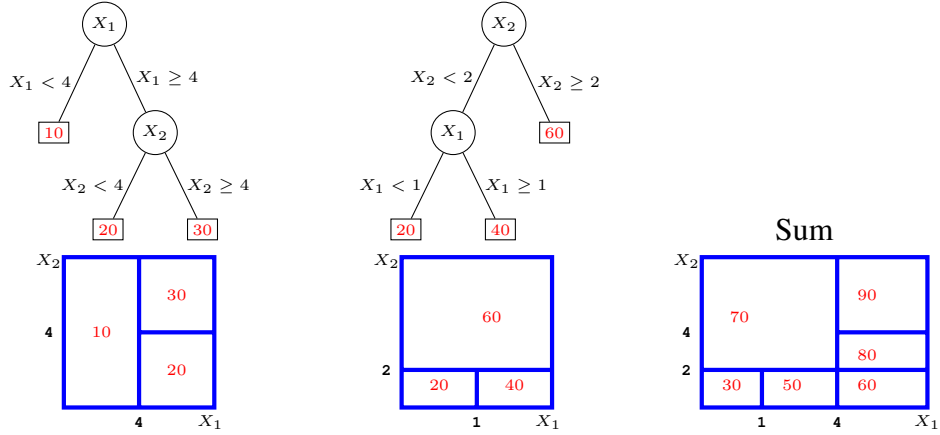
\begin{figure}[t]
\centering
\medskip
{ \tiny
\hspace{-2cm}
  \begin{forest}
  for tree={font=\tiny,l sep =4em, s sep=4em, anchor=center} 
    [$X_1$, circle, draw,
       [\textcolor{red}{$10$},rectangle,draw, edge label={node[midway,left]{$X_1<4$}}]
       [$X_2$, edge label={node[midway,right]{$X_1\ge 4$}}, circle, draw,  
         [\textcolor{red}{$20$}, rectangle,draw,edge label={node[midway,left]{$X_2<4$}}]
         [ \textcolor{red}{$30$}, rectangle,draw,edge label={node[midway,right]{$X_2\ge 4$}}]]]
  \end{forest}
 }
  { \tiny
  \hspace{1cm}
  \begin{forest}
  for tree={font=\tiny,l sep =4em, s sep=4em, anchor=center} 
    [$X_2$, circle, draw,
     [$X_1$, edge label={node[midway,left]{$X_2< 2$}}, circle, draw,  
         [\textcolor{red}{$20$}, rectangle,draw,edge label={node[midway,left]{$X_1<1$}}]
         [ \textcolor{red}{$40$}, rectangle,draw,edge label={node[midway,right]{$X_1\ge 1$}}]]
         [\textcolor{red}{$60$},rectangle,draw, edge label={node[midway,right]{$X_2\ge 2$}}]]
  \end{forest}
 }
\hspace{2cm}
{Sum}
 
 \hspace{-0.2cm}
\begin{tikzpicture}[num/.style={font=\tiny\bfseries\ttfamily}]
 \draw[ultra thick,blue] (0,0) rectangle (2,2)
                        (1,0) -- (1,2)
                        (1,1) -- (2,1); 
 \node[num] at (1,-0.2){4};
 \node[num] at (-0.2,1){4} ;
 \node[num] at (1.95,-0.2){$X_1$};
 \node[num] at (-0.2,1.95){$X_2$} ;  
 \node[num] at (0.5,1){\textcolor{red}{$10$}} ;
 \node[num] at (1.5,0.5){\textcolor{red}{$20$}} ;
 \node[num] at (1.5,1.5){\textcolor{red}{$30$}} ;
 \end{tikzpicture}
 \hspace{1.5cm}
\begin{tikzpicture}[num/.style={font=\tiny\bfseries\ttfamily}]
 \draw[ultra thick,blue] (0,0) rectangle (2,2)
                        (0,0.5) -- (2,0.5)
                        (1,0) -- (1,0.5); 
 \node[num] at (1,-0.2){1};
 \node[num] at (-0.2,0.5){2} ;
 \node[num] at (1.95,-0.2){$X_1$};
 \node[num] at (-0.2,1.95){$X_2$} ;  
 \node[num] at (0.5,0.3){\textcolor{red}{$20$}} ;
 \node[num] at (1.5,0.3){\textcolor{red}{$40$}} ;
 \node[num] at (1.2,1.2){\textcolor{red}{$60$}} ;
 \end{tikzpicture} 
 \hspace{1cm}
\begin{tikzpicture}[num/.style={font=\tiny\bfseries\ttfamily}]
 \draw[ultra thick,blue] (0,0) rectangle (3,2)
                        (0,0.5) -- (3,0.5)
                        (0.8,0) -- (0.8,0.5)
                        (1.8,0) -- (1.8,2)
                        (1.8,1) -- (3,1); 
\node[num] at (1.8,-0.2){4};
 \node[num] at (-0.2,1){4}; 
 \node[num] at (0.8,-0.2){1};
 \node[num] at (-0.2,0.5){2} ;
 \node[num] at (3,-0.2){$X_1$};
 \node[num] at (-0.2,1.95){$X_2$} ;  
 
 \node[num] at (0.4,0.3){\textcolor{red}{$30$}} ;
 \node[num] at (0.5,1.2){\textcolor{red}{$70 $}} ;
 \node[num] at (2.2,0.7){\textcolor{red}{$80$}} ;
 \node[num] at (1.2,0.3){\textcolor{red}{$50$}}; 
 \node[num] at (2.2,0.3){\textcolor{red}{$60$}} ;
 \node[num] at (2.2,1.5){\textcolor{red}{$90$}} ;
 \end{tikzpicture}
  \caption{Two trees and their sum}
  \label{fig:trees}
\end{figure}

Consider the problem of making inference about a function $f$ that predicts a continuous outcome $Y$ using a vector of predictors $x=(x_1,\cdots,x_p)$ given that  
\[
 Y = f(x) +\epsilon, 
   \quad \quad  \epsilon \sim N(0,\sigma^2)
\]
One way to solve this inference problem is to approximate $f(x)= E(Y|x)$ by a nonparametric function. With BART, $f(x)$ is approximated by a sum-of-trees model;

\begin{equation}\label{bart}
 Y = \sum_{j=1}^{m} g(x;T_j,M_j) + \epsilon, 
    \quad \quad  \epsilon \sim N(0,\sigma^2)
\end{equation}
where $T_j$ is a binary tree, $M_j = \{ \mu_{1j},\mu_{2j},\cdots, \mu_{Bj}\}$ the $j^{th}$ tree's terminal nodes parameters, and $g(x;T_j,M_j)$ a tree function that assigns a $\mu_{bj} \in M_j$ to $x$ as illustrated in Figure \ref{fig:trees}. Under equation (\ref{bart}), $E(Y|x)$ is then calculated as the sum of all terminal nodes, $\sum_{j=1}^m\mu_{bj}$, assigned to $x$ by each tree function $g(x;T_j,M_j)$, where $b = 1,\cdots,B$. Figure \ref{fig:trees} depicts such a sum. Chipman \textit{et al.} \citep{bart2010} showed the great predictive ability of BART compared to other ensemble menthods via simulations and applied examples and developed a BART R package to easily fit this model \citep{bartmanual}. 

The independent prior for the error variance is $\sigma^2 \sim \nu\lambda \chi^{-2}(\nu)$.  Notationally, we specify the prior for the unknown function, $f$, as $f\sim BART$.  
This prior is made up of two components: a prior for the tree structures, $T_j$ and a prior for the terminal nodes parameters, $M_j|T_j$, so that  
\[
p((T_1,M_1),\cdots,(T_m,M_m)) = [\prod_j p(M_j|T_j)p(T_j)]
\]
 
The prior for $T_j$ is specified by describing a probabilistic process for growing a tree, where P(node at depth d is non-terminal)= $\alpha (1+d)^{\beta}$ , $\alpha \in (0,1)$ and $\beta \in [0,\infty)$. Using default values of $\alpha=0.95, \beta = 2$, trees with 2 or 3 terminal nodes are favored as shown in table \ref{tabtree}. At each internal node, a splitting variable and splitting value are chosen uniformly among all the available variables and cutoff values, respectively. The terminal nodes are given independent normal priors, $\mu_{bj} \sim N(0, \sigma^{2}_{\mu})$ with $\sigma_{\mu} = 0.5/k\sqrt{m}$, where $k$ is a tuning parameter with default value of $2$.
 
\begin{table}[H]
\begin{center}
\begin{tabular}{|c|c  c c c c|}
 \hline
 Number of terminal nodes, B & 1  & 2 &  3 & 4 & 5+ \\
 \hline
Prior probability & 0.05  & 0.55  & 0.28  & 0.09 & 0.03\\
\hline
\end{tabular}
\caption{Tree complexity given default prior settings: $\alpha=0.95, \beta = 2$ ~\citep{SparLoga16},\label{tabtree}}
\end{center}
\end{table}  

\subsection{Bayesian Additive Regression Trees (BART); binary response}
\label{sec:bart_bin}

In clinical research, studies with binary endpoints arise often. Chipman \textit{et al.} \citep{bart2010} provide a detailed extension of the above BART model to handle binary endpoints using a probit link and the data augementation idea of Albert and Chib \citep{AlbeChib93}. However, this model does not directly provide the covariate effects in terms of odds ratios. The current BART R package \citep{bartmanual} already includes an implementation of the BART model for binary outcomes using a logit link and data augumentation approach of Holmes and Held \citep{HolmHeld06}. Consider a binary outcome $Y= 0,1$, then the logit model set up for classification is 
\begin{align*}\label{lbart}
  p(x) & \equiv P[Y=1|x] = F(f(x)), \text{ where }    \\
   f(x) & \equiv \sum_{i=1}^{m} g(x;T_j,M_j)
\end{align*}
Here, $F$ represents the standard Logistic cumulative distribution function.
The Holmes and Held augmentation idea introduces latent variables $z_1,\cdots,z_n$ such that:

\begin{align*}
  y_i        & = \begin{cases}
                  1 & \text{ if}  z_i > 0 \\
                  0 & \text{ Otherwise}
                 \end{cases}\\
  z_i        & =  f(x) + \epsilon_i \\
  f          & \sim BART \\
  \epsilon_i & \sim N(0,\lambda_i)\\
  \lambda_i & = (2\phi_i)^2 \\
  \phi_i    & \sim KS\ .
\end{align*}
The prior hyperparameter for the terminal node parameters is modified to $\sigma_{\mu} = 3/k\sqrt{2m}$, but the prior for $f$ is otherwise unchanged.  Inferences for $f$ are available using Gibbs sampler for draws $[z|f,\lambda]$ using a truncated normal distribution as described in \citep{HolmHeld06}, $[f|z,\lambda]$ using the approach detailed in \citep{bart2010} for parameters in the sum-of-trees model  with $z$ as the outcome and $[\lambda|f,z]$ following the rejection sampling algorithm detailed in \citep{HolmHeld06}.

\subsection{Log Pseudo Marginal Likelihood (LPML)}
\label{sec:lpml}

In Bayesian model comparison, Bayes factors \citep{KassRaft1995,HanCarl2001,Gelf1995,ChrisJohn11} have been used as a measure of evidence in favor of one model as opposed to another, and clear threshold values that indicate the strength of this evidence have been established \citep{jeffreys1961, KassRaft1995}. Assume that the data $y_1,\cdots,y_n$ are conditionally independent given a model $\mathcal{M}$ and model parameters $\theta_M$. Then the marginal likelihood is given by :
\begin{equation*}
p(\mathbf{y}|\mathcal{M}) = \int_{\theta_{\mathcal{M}}}\prod_{i=1}^{n} p(y_i|\theta_{\mathcal{M}},\mathcal{M}) p(\theta_{\mathcal{M}}) d\theta_{\mathcal{M}}\ .
\end{equation*}
Bayes factors are calculated as the ratio of the marginal likelihoods from two competing models. They can be interpreted as the ratio of posterior odds to prior odds for the model in the numerator of the Bayes factor.  However, $p(\mathbf{y}|\mathcal{M})$ is not available analytically for improper or even vague priors. 

Geisser and Eddy \citep{GeisEddy79} suggested replacing $p(\mathbf{y}|\mathcal{M})$  by the pseudo marginal likelihood (PML)

\[
\tilde{p}(\mathbf{y}|\mathcal{M}) = \prod_{i=1}^{n} p(y_i|y_{-i},\mathcal{M})
\]
where $p(y_i|y_{-i},\mathcal{M})$ is the $i^{th}$ conditional predictive ordinate (CPO$_i$). This CPO$_i$ is the predictive density calculated at the observed $y_i$ given all data except the $i^{th}$ observation. Thus, the log pseudo marginal likelihood (LPML) is computed as 
\[ 
  LPML = \sum_{i=1}^{n} \log(CPO_i)\ .
\]
Given two competing models, a model that maximizes the LPML is chosen. Moreover, a pseudo Bayes factor (PsBF) is calculated by exponentiation of the difference between LPML statistics of the two competing models, i.e., PsBF is a ratio of pseudo marginal likelihoods. Threshold values as those in \citep{jeffreys1961} are used to choose the data supported model. 

LPML and PsBF gained widespread use \citep{DeyChen97, ChenIbra02, OMalNorm03, ChanGian04, LiBolt06, ChanGian06, HansTim2007, BhatSeng09, HansTim2011, HeydFu16, ChenGill17} after Gelfand and Dey \citep{GelfandDey1994} established asymptotic properties of the PML and showed that CPO$_i$ can easily be estimated using posterior samples $\theta_M^1,\cdots,\theta_M^S$ by
\[
 \hat{CPO}_i = \frac{1}{S} \sum_{s=1}^{S} \frac{1}{\tilde{p}(y_i|\theta_{\mathcal{M}}^s,\mathcal{M})}\ .
\] 	
Here $\tilde{p}$ represents the probability density function (pdf) if the outcome is continuous and represents the probability mass function for a binary outcome. Note that this is readily available for BART models for both continuous and binary response given $f(\cdot)$ (and $\sigma$ in the former case) simply as the normal density and the Bernoulli mass function.

\subsection{Out-of-sample prediction error (OSPE)}
\label{sec:ospe}
 The out-of-sample prediction error (OSPE), based on 5-fold cross-validation,  is another criterion commonly used in assessing a model's predictive performance. The dataset is randomly split into 5 equal (or nearly equal) sized folds. For each $k=1,\cdots,5$, a training dataset is constructed using all data except those in the k$^{th}$ fold, and a test data set using data points from the k$^{th}$ fold. The model is fit to the training data, and used to evaluate the prediction error using the test dataset.  Finally, one averages these fold-based errors to obtain the OSPE. Thus, the OSPE is calculated as
\[
  OSPE = \frac{1}{n}\sum_{k=1}^{5}\left( \sum_{i=1}^{n_k}(Y_i - \hat{Y}_i)^2\right)\ ,
\]
where $Y_i$ is i$^{th}$ observation from the test dataset and $\hat{Y_i}$ the fitted value. For competing models, the model with the lowest OSPE is preferred. A ratio of OSPEs from two competing models is sometimes used for quantifying model comparison.

\section{Additive nonparametric models}  
\label{sec:add}
	
BART models described in sections \ref{sec:bart} and \ref{sec:bart_bin} are known to work well when prediction is the primary goal. In addition to predictions, researchers are usually interested in the interpretation of at least some aspects of the fitted model, e.g., individual covariate effects. The predictive function from BART is not available analytically and requires post processing to recover aspects such as individual covariate effects. While this is possible and often undertaken, to further assist in interpretation we introduce a sum of two simpler BART models which we refer to as the additive BARTs model.
	
\subsection{ Additive BARTs model; continuous outcome}
\label{twobarts}

Consider the additive BARTs model written as:
\begin{align*}
 Y&= f_1(x^{-}) + f_2(x^{+}) + \epsilon\\ 
    &= \sum_{j=1}^{m} g^{-}(x^{-},T^{-}_{j},M^{-}_{j}) + \sum_{j=1}^{m} g^{+}(x^{+},T^{+}_{j},M^{+}_{j}) + \epsilon , \quad \quad \quad \epsilon \sim N (0,\sigma^2)
\end{align*} 
where $x^{+}$, $x^{-}$ are two disjoint sets of the covariates and $f_1()$, $f_2()$ are the flexible functions fit with $x^{-}$, $x^{+}$ , respectively. Each function $f_k$ can be written as a sum of trees as in the original BART model. The parameters of this additive BARTs model are $(T^{-}_{1},M^{-}_{1},\cdots, T^{-}_{m},M^{-}_{m},T^{+}_{1},M^{+}_{1},\cdots, T^{+}_{m},M^{+}_{m},\sigma^{2} )$.
 We place BART priors on $f_1(.)$ and $f_2(.)$; i.e, $f_1 \sim BART$ and $f_2 \sim BART$. We use the same prior specification as for a single BART model, except that the terminal nodes are given a normal prior , $\mu_{ij} \sim N(0, \sigma^{2}_{\mu})$ with 
\[ 
\begin{cases}
\sigma_{\mu} = 0.5/k\sqrt{2m} & \text{ continuous outcome}\\
\sigma_{\mu} = 3/k\sqrt{2m}   & \text{ Binary outcome }\ .
\end{cases}
\]
 We add to the specification in \citep{bart2010} this factor of 2 in front of $m$ so that the additive effect of all the trees in the two additive BARTs model remains stochastically equivalent to a corresponding single ensemble model of $m$ trees.  

To make inference based on this additive BARTs model, we use the Blocked Gibbs Sampling technique \citep{Gelf1995} to sample from the posterior distribution $[(T^{-}_{1},M^{-}_{1},\cdots, T^{-}_{m},M^{-}_{m},T^{+}_{1},M^{+}_{1},\cdots, T^{+}_{m},M^{+}_{m},\sigma^{2} )|y]$. Define $M^{-} = M^{-}_{1},\cdots,M^{-}_{j}$, $T^{-} = T^{-}_{1},\cdots,T^{-}_{j}$, $M^{+} = M^{+}_{1},\cdots,M^{+}_{j}$, and $T^{+} = T^{+}_{1},\cdots,T^{+}_{j}$.  This algorithm first draws samples for $ [T^{+},M^{+}| T^{-},M^{-},\sigma^{2},y]$, then samples for $[T^{-},M^{-}| T^{+},M^{+},\sigma^{2},y]$, and finally samples of $ [\sigma^{2}|T^{-},M^{-},T^{+},M^{+},y] $ from their posterior distributions. To accomplish this, consider the residuals:
\begin{align*}
Y^{-} &= Y-\sum_{j=1}^{m} g^{-}(x^{-},T^{-}_{j},M^{-}_{j})= \sum_{j=1}^{m} g^{+}(x^{+},T^{+}_{j},M^{+}_{j}) + \epsilon\\
Y^{+} &= Y -\sum_{j=1}^{m} g^{+}(x^{+},T^{+}_{j},M^{+}_{j})= \sum_{j=1}^{m} g^{-}(x^{-},T^{-}_{j},M^{-}_{j}) + \epsilon
\end{align*}
Notice that ($T^{+}_{j},M^{+}_{j}$) depends on ($T^{-}_{j},M^{-}_{j}$) only through the residuals $Y^{-}$, and vice versa. Hence, the Blocked Gibbs sampling algorithm first draws samples for $[T^{+},M^{+} |\sigma^{2},Y^{-}]$ as described in \citep{bart2010} for a single BART model applied to $Y^{-}$. In a similar manner, we make draws for $[T^{-},M^{-}|\sigma^{2},Y^{+}]$ using $Y^{+}$. Given all the trees and terminal nodel parameters, we then obtain draws for $[\sigma^{2}|T^{-},M^{-},T^{+},M^{+},Y^{+},Y^{-}]$ from an inverse gamma distribution.\\

To get draws for the first step of this algorithm,  we use the property described in \citep{bart2010} that shows that the conditional distributions $ p(T^{-}_{j},M^{-}_{j}| T^{-}_{(j)},M^{-}_{(j)},\sigma^{2},Y^{-})$ depend on $T^{-}_{(j)},M^{-}_{(j)}$ only through the residuals
  \[
  R^{-}_{j} = Y^{-}-\sum_{k\ne j} g^{+}(x^{+},T^{+}_{k},M^{+}_{k}) \ .
  \] 
Similary for step 2, we get the residuals 
  \[ 
  R^{+}_{j} = Y^{+}-\sum_{k\ne j} g^{-}(x^{-},T^{-}_{k},M^{-}_{k}) \ .
  \] 
Thus we have the steps:  \\
\textbf{Step 1:} $[T^{-}_{j},M^{-}_{j}| R^{-}_{j} ,\sigma^{2} ]$\\
\textbf{Step 2:} $ [T^{+}_{j},M^{+}_{j}| R^{+}_{j},\sigma^{2} ]$\\
\textbf{Step 3:} $ [\sigma^{2}|T^{-},M^{-},T^{+},M^{+},y]$.
 
\noindent
In step 1, samples for $T^{-}_{j}| R^{-}_{j} ,\sigma^{2} $ are drawn using the Metropolis-Hastings algorithm as described in \citep{bart2010} and $M^{-}_{j}| R^{-}_{j} ,\sigma^{2}$ is made of independent draws of the terminal nodes parameters $\mu_{ij}$ from a normal distribution. In a similar manner, we draw parameters from step 2. Finally, $\sigma^{2}|T^{-},M^{-},T^{+},M^{+},y$ is drawn from an inverse gamma distribution. The above algorithm produces a sequence of draws  $(T^{-}_{1},M^{-}_{1}),\cdots, (T^{-}_{m},M^{-}_{m}),(T^{+}_{1},M^{+}_{1}),\cdots, (T^{+}_{m},M^{+}_{m}),\sigma^{2}$ which converges in distribution to the posterior distribution $p((T^{-}_{1},M^{-}_{1}),\cdots, (T^{-}_{m},M^{-}_{m}),(T^{+}_{1},M^{+}_{1}),\cdots, (T^{+}_{m},M^{+}_{m}),\sigma^{2}|y)$. The resulting sequence of sum-of-trees functions  
\[
f^*(x) = \sum_{j=1}^{m} g^{-}(x^{-},T^{*-}_{j},M^{*-}_{j}) + \sum_{j=1}^{m} g^{+}(x^{+},T^{*+}_{j},M^{*+}_{j}) 
\]
converges to the true posterior distribution $p(f|y)$. To obtain the predicted value of Y at a given value of the covariate vector $x$, we use the average of the after burn-in sample 
\[
\frac{1}{S}\sum_{s=1}^{S} f^{*}_{s}(x)
\]
which is an approximation of the posterior mean $ E(f(x)|y)$ with $S$ being the number of Gibbs samples after burn-in.

\subsection{Additive treatment BART model; continuous outcome}
\label{onebt}

The additive BARTs model introduced in section \ref{twobarts} is general in its specification and does not directly provide  estimates and interpretation of individual covariate effects. If we are interested in the estimation and interpretation in the effect of a treatment, however, an alternative additive model is readily available almost as a special case of the two BARTs model. To elaborate, we propose fitting a BART model to the set of covariates and a parametric model for the treatment. This model allows for detection of interaction between the treatment and the set of covariates. A model without interaction between the treatment and other variables is less cumbersome and has a directly interpretable treatment effect. To specify this model, let $A_i$ be the treatment variable and $x_i$ a vector of other covariates. Then, define the additive treatment BART model as:
\begin{align*}
y_i &= \beta A_i  + f(x_i) + \epsilon_i\\ 
    &=  \beta A_i + \sum_{j=1}^{m} g^{+}(x_i,T^{+}_{j},M^{+}_{j}) + \epsilon_i 
\end{align*}
with  $\epsilon _i \sim N (0,\sigma^2)$ and $i=1,\cdots,n$.
Then the likelihood for this model is given by: 
\[ 
[y_1,\cdots,y_n|\beta,\sigma^{2},f] = \frac{1}{\sqrt{2\pi\sigma^2}}
                                    e^{-\frac{1}{2\sigma^2}\sum_{i=1}^{n}[ y_i -(\beta A_i + f(x_i))]^2}
\]
We assume that the prior distribution for the treatment effect $\beta$  is $N(\mu_{0},\sigma^{2}_{0})$. By using the standard conjugacy argument, the posterior distribution for $\beta$ is $N(\hat{\mu}_\beta, \hat{\sigma}^{2}_{\beta})$ with 
\begin{align*}
\hat{\mu}_\beta &= \frac{\sum_{i=1}^{n}( y_i  - f(x_i))A_i\sigma^{2}_{0}+\mu_0 \sigma^2}{\sum_{i=1}^{n} A_{i}^{2}\sigma^{2}_{0} +\sigma^2}\\ 
\hat{\sigma}^{2}_{\beta} &= \frac{\sigma^2 \sigma^{2}_{0}}{\sum_{i=1}^{n} A_{i}^{2}\sigma^{2}_{0} +\sigma^2}\ .
\end{align*}

To get posterior samples for the parameters in this model, we follow a similar Blocked Gibbs Sampling algorithm as that outlined in section \ref{twobarts}. We first get samples for  $[T^{+},M^{+}|\sigma^2,y^{-}]$ using $y^{-} = y-\beta A$ followed by sampling $[\beta|y^{-}]$ from its normal posterior distribution using the residuals $y^{+} = y-\sum_{j=1}^{m} g^{+}(x,T^{+}_{j},M^{+}_{j})$. Finally, we sample for $[\sigma^2|T^{+},M^{+},y^{+},y^{-}]$ from an inverse gamma distribution.

\subsection{ Additive BARTs model; binary outcome}
\label{twobarts:binary}

Akin to Section \ref{twobarts}, we introduce an additive BART model for a binary response to assist in interpretation.  Consider a binary outcome $Y= 0,1$, then the additive logit model set up is 
\begin{align*}
  p(x) \equiv P[Y=1|x] &  = F( f_1(x^{-}) + f_2(x^{+}) )  \\
                        & =  F(\sum_{j=1}^{m} g^{-}(x^{-},T^{-}_{j},M^{-}_{j}) + \sum_{j=1}^{m} g^{+}(x^{+},T^{+}_{j},M^{+}_{j})) \ .
\end{align*}
Here, $F$ represents the standard Logistic cumulative distribution function. Again taking the Holmes and Held augmentation idea \citep{HolmHeld06}, we use latent variables $z_1,\cdots,z_n$ as:

\begin{align*}
  y_i        & = \begin{cases}
                  1 & \text{ if }  z_i > 0 \\
                  0 & \text{ Otherwise}
                 \end{cases}\\
  z_i        & = f_1(x^{-}) + f_2(x^{+})+ \epsilon_i \\
  f_1  &\sim \text{ BART},f_2  \sim \text{ BART}\\
  \epsilon_i & \sim N(0,\lambda_i)\\
  \lambda_i & = (2\phi_i)^2 \\
  \phi_i    & \sim KS \ . \\
\end{align*}
In order to make inference about this additive model, we need samples of $z|f_1,f_2,\lambda,\beta$; $f_1|z,f_2,\lambda$; $f_2|f_1,z,\lambda$ and $\lambda|f_1,f_2,z$. Consider the residuals $z^{-}_{i} = z_i - f_1(x^{-}) = f_2(X^{+}) + \epsilon_i$ and  $z^{+}_{i} = z_i - f_2(x^{+}) = f_1(X^{-}) + \epsilon_i$. Notice that $f_1$ depends on $f_2$ only through the residuals $z^{+}$. The rest of the steps for the Blocked Gibbs sampling algorithm follow a similar approach as that described in Section \ref{twobarts} to get samples of $z|f_1,f_2,\lambda,\beta$; $f_1|z^{+},f_2,\lambda$; $f_2|z^{-},f_1,\lambda$ and $\lambda|f_1,f_2$. Holmes and Held \citep{HolmHeld06} provide an algorithm to get samples of $\lambda|f_1,f_2$. Samples of  $z|f_1,f_2,\lambda,\beta$ are obtained using a truncated normal sampler.

\subsection{Additive treatment BART model; binary outcome}
\label{onebt:binary}
 
Here we modify the model introduced in Section \ref{twobarts:binary} by replacing the specification for $f_1$ by a simple parametric form for the treatment effect. Consequently, notationally replacing $f_2$ by $f$ as there is only one nonparametric function, the changes to the two BARTS binary model are:
\begin{align*}
  z_i        & =  \beta A_i  + f(x_i) + \epsilon_i \\
  \beta      & \sim N(\mu_{0},\sigma^{2}_{0}) \\
\end{align*}
where $z_i$'s are latents as in Section \ref{twobarts:binary}. Posterior sampling also proceeds as there except that $[\beta|z^{+}_{i},\lambda] \sim N(B,V)$ with
\begin{align*}
V &=  [(\sigma_{0}^{2})^{-1} + A^{'}diag(\lambda_1^{-1},\cdots,\lambda_n^{-1})A]^{-1}\\
B &= V ((\sigma_{0}^{2})^{-1} \mu_0 + A^{'}diag(\lambda_1^{-1},\cdots,\lambda_n^{-1})z^{+}) \ .
\end{align*}

\subsection{Using LPML and OSPE to assess nonadditivity}

Considering the ease with which LPML is calculated given the posterior samples, we propose using the LPML criterion described in section \ref{sec:lpml} in conjunction with all the BART models described above to assess whether a model is additive or not. To accomplish this, we first fit data to single BART model (nonadditive), and obtain $LPML_1$. Next, we fit an additive BARTs model from Sections \ref{twobarts} or \ref{onebt} (Sections \ref{twobarts:binary} or \ref{onebt:binary} if the outcome is binary) with the covariate space divided into two disjoint sets of covariates, and obtain $LPML_2$. Given $LPML_1$ and $LPML_2$, we calculate the  Pseudo Bayes Factor as $PsBF = e^{LPML_1 - LPML_2}$. We interpret the PsBF following the guidelines proposed by Jeffreys \citep{jeffreys1961} shown in table \ref{tab2} ; i.e, if the PsBF is above 3, we conclude that there is substantial evidence in favor of the nonadditive model, and so on. Given the OSPE$_1$ from nonadditive and OSPE$_2$ from additive BARTs models, we calculate $R_{OSPE} = \frac{OSPE_1}{OSPE_2}$. If $R_{OSPE}$ is above 1, we conclude that the model is additive. The existence of these threshold values for PsBF provides an important advantage over other criteria such as AIC or OSPE which do not provide clear graded thresholds for model selection. Moreover, given the cross-validation required in computing OSPE, it is computationally expensive compared to PsBF which is very fast to calculate.
\begin{table}[H]
 	\begin{center}
 	  \begin{tabular}{|c|c|}
 	  \hline
 	   BF (PsBF) & Strength of evidence\\
 	   	  \hline
 	   $< \frac{1}{3}$ & Evidence for $H_0$\\
 	   $\frac{1}{3}$  to 3 & Barely worth mentioning \\
 	   3 to 10 & Substantial\\
 	   10 to 30 & Strong \\
 	   30 to 100 & Very strong \\
 	   $>$ 100 & Decisive \\
 	   \hline
 	  \end{tabular}
 	 \caption{Threshold values for interpreting BFs and PsBFs proposed by Jeffreys \citep{jeffreys1961}}
 	 \label{tab2}
 	\end{center}
 \end{table}

When these additivity assessments compare one model with a single BART and another with two additive BARTs, we recommend that in the latter model each BART use half the number of trees used in the single BART model. This maintains comparability of the priors for the two models. We followed this recommendation in all calculations in this article. 
 
\section{Designing simulations to assess additivity}
\label{sim:design}

In the next section, we use simulations to demonstrate the performance of LPML and OSPE criteria to assess additivity in BARTs models. However, the variability of each term in the model relative to others affects whether additivity can be detected or not. Moreover, large random error term can cause the effect of other terms in the model to be undetectable. Hence, we propose a systematic approach to design simulations in which the variability of each term is standardized relative to others.
 
\subsection{Simulation design for a continuous outcome}
\label{des:cont}
Consider generating data from a nonadditive model with continuous outcome $Y$ and disjoint sets of covariates $X^{-}$ and $X^{+}$, where we express

\[
Y = g(X^{-}) + c \beta_1 f(X^{+}) + \beta_2 f(X^{+}) h(X^{-}) +\epsilon, \epsilon \sim N(0, \sigma^2)
\]
with $X^{-}, X^{+},\epsilon$ jointly independent. For a corresponding additive model, we define  $Y = g(X^{-}) + \beta_1 f(X^{+}) +\epsilon, \epsilon \sim N(0, \sigma^2)$; i.e, we assume $c = 1$ and $\beta_2 = 0$. 

A major factor in determining the degree of nonadditivity is the variability in each of the terms of this model relative to the total variance function. We propose an approach that standardizes the variance of each term relative to the total variance function. To simplify the notation, we use $h_X^{-} = h(X^{-}), f_X^{+} = f(X^{+})$, and $g_X^{-} = g(X^{-})$. Then the variance function is given by:
\begin{align*}
var(Y) =& var(g_{X^{-}}) + var(c \beta_1 f_X^{+}) + var(\beta_2 f_X^{+} h_X^{-}) + \sigma^2 \\
      & +   2 Cov(g_X^{-}, \beta_2 f_X^{+} h_X^{-}) + 2 Cov(\beta_1 f_X^{+} , \beta_2 f_X^{+} h_X^{-}) 
\end{align*}  
 
We propose the standardization ratios $\alpha$, $\delta$, $\gamma$ where $\alpha$ is the variability attributed to random noise, 
$\delta$ the variability due to both main and interaction effects of $X^+$ variables, and $\gamma$ the variability attributed to the interaction between $X^{-}$ and $X^{+}$ variables.  For additive models with no interaction between the two sets of variables, $\delta$ represents the variability due to the main effects of $X^+$ variables. However, for models with interactions between $X^{-}$ and $X^{+}$, $\delta$ represents the variability due to both the main and interaction effects of $X^+$ variables. Hence, our standardization ratios are defined as:
\begin{align}
\alpha       &= \frac{\sigma^2}{var(Y)} \label{eq:alpha}\\
 \delta &= \frac{var(c\beta_1 f_X^{+} + \beta_2 f_X^{+} h_X^{-})}{var(g_X^{-} + c\beta_1 f_X^{+} + \beta_2 f_X^{+} h_X^{-})}\label{eq:delta}\\
\gamma       &= \frac{var(\beta_2 f_X^{+} h_X^{-})}{var(g_X^{-} + c\beta_1 f_X^{+} + \beta_2 f_X^{+} h_X^{-})}\ . \label{eq:gamma} 
\end{align}
Setting fixed values of $\alpha, \delta, \gamma$, we obtain a system of quadratic equations in $\beta_1, c $, and $\beta_2$ and solve them analytically to get appropriate sulutions for $\sigma$, $\beta_1, c $, and $\beta_2$.   
  
\subsection{Simulation design for a binary outcome}
\label{des:bin}	
 We generate $g_0(X^{-}), f_0(X^{+})$ and $h_0(X^{-})$ with $X^{-}$ and $X^{+}$  generated independently and center the above functions by: $g(X^{-}) = g_0(X^{-}) -\overline{g_0(X^{-})}$, $f(X^{+}) = f_0(X^{+}) -\overline{f_0(X^{+})}$, and $h(X^{-}) = h_0(X^{-}) -\overline{h_0(X^{-})}$.  This centering is used to ensure thefunction values are located in a reasonable range on the logit scale.
 
Consider data generated from the following model with an interaction between $X^{-}$ and $X^{+}$:
\[
l(Y) = \beta_0 g(X^{-}) + c \beta_1 f(X^{+}) + \beta_2 f(X^{+}) h(X^{-})
\]
with $X^{-}, X^{+}$ jointly independent, and $l(Y)$ being a link function such as the logit or probit function.\\
 Let $h_X^{-} = h(X^{-}), f_X^{+} = f(X^{+})$, and $g_X^{-} = g(X^{-})$. Then 
\begin{align*}
var(l(Y))= & \beta_0 ^2 var(g_X^{-}) + var(c \beta_1 f_X^{+}) + var(\beta_2 f_X^{+} h_X^{-})  \\
&+ 2 Cov(\beta_0g_X^{-}, \beta_2 f_X^{+} h_X^{-}) + 2 Cov(\beta_1 f_X^{+} , \beta_2 f_X^{+} h_X^{-}) 
\end{align*}
 
Motivated by prior specifications outlined in  \citep{bart2010} concerning BART probit for classification, let 
\[var(\beta_0 g(X^{-}) + c \beta_1 f(X^{+}) + \beta_2 f(X^{+}) h(X^{-}))= \nu =1.5^2\]
 Following a similar approach as that described in Section \ref{des:cont} to standardize the variance of each term relative to the total variance, we define the following normalization ratio $\delta$ that represents the variability due to the effects of $X^+$, 
  
\begin{align}
\delta &= \frac{var(c\beta_1 f_X^{+} + \beta_2 f_X^{+} h_X^{-})}{var(\beta_0 g_X^{-} + c\beta_1 f_X^{+} + \beta_2 f_X^{+} h_X^{-})} \label{eq:deltab} \\
&= \frac{var(c\beta_1 f_X^{+}) + var(\beta_2 f_X^{+} h_X^{-})+2Cov(c\beta_1 f_X^{+},\beta_2 f_X^{+} h_X^{-})}{\nu} \nonumber  
\end{align} 
and $\gamma$ the variability due to interaction between the two sets of variables: 
\vspace{-0.1cm}
\begin{align}
\gamma &= \frac{var(\beta_2 f_X^{+} h_X^{-})}{ var(\beta_0g_X^{-} + c\beta_1 f_X^{+} + \beta_2 f_X^{+} h_X^{-})} \label{eq:gammab} \\
&= \frac{\beta_2^2 var(f_X^{+} h_X^{-})}{\nu}  \nonumber 
\end{align}
Forcing $\beta_1$ to be the same for both additive and non-additive models and setting fixed values of $\delta$ and $\gamma$, we analytically solve equations (\ref{eq:gXb}), (\ref{eq:deltab}),(\ref{eq:gammab}) to get appropriate values of  $\beta_0, \beta_1, c $, and $\beta_2$.\\

\section{Simulation Settings and Results}
\label{sims}
We use the systematic approach of Section \ref{sim:design} to conduct simulations tailored for additivity testing, and compare performance of LPML and OPSE in detecting additivity. For continuous outcomes, we generate data using 
\[
Y = g(X^{-}) + c\beta_1 f(X^{+}) + \beta_2 f(X^{+}) h(X^{-}) + \epsilon,  \text{with } \epsilon \sim N(0, \sigma^2)\ .
\]
For binary data, we use 
 \[
l(Y) = \beta_0 g(X^{-}) + c \beta_1 f(X^{+}) + \beta_2 f(X^{+}) h(X^{-})\ .
\]
Choices of $f,g,h$ are described in the next subsection along with settings of $\alpha,\delta,\gamma$. The second subsection describes simulation results.
 
\subsection{Simulation Settings}

We consider four different scenarios for our simulations as shown by the different choices of $g$ and $h$ given in Table \ref{tab:settings}. The proposed scenarios vary by degree of complexity, from simple linear functions in Scenario 1 to nonlinear functions in Scenario 4. For the choice of $f$, we use a commom $f(X^{+})= (X_6 + X_7)^2$ with $ X^{+} = X_6,X_7 \stackrel{ind}{\sim}$  $N(0,1)$ in assessing additivity. In the additive treatment case, we use $A \sim Bernoulli(0.5)$ representing a binary treatment.   

\begin{table}[b]
\begin{adjustbox}{max width=\textwidth}
\begin{tabular}{|p{1.5cm}|p{8cm}|p{5.8cm}|p{6.9cm}|}
\hline  
\bf{Scenario} & \bf{$g(X^{-})$} &\bf{$h(X^{-})$} & \bf{Covariates distribution}\\
\hline
1. & $3 + 3X_1-3X_2-2X_3 $ & $-2X_1-7X_2-X_3 + 3X_2X_3$ & $X_1,X_2, X_3 \stackrel{ind}{\sim}  N(0,1)$ \\
\hline
2. & $2-3X_{1}^{2} -3X_{2}^{2}+ 3X_1X_2$ & $-X_1+X_2 -2.5X_1X_2$ & $X_1,X_2 \stackrel{ind}{\sim}U(-3,3)$\\
\hline
3. & $ 3.5 - X_1 + X_2 + 2I(X_3<1)$ & $  X_1-0.5X_2-3I(X_3<1)$ & $X_1,X_2, X_3 \stackrel{ind}{\sim} N(0,1)$\\
\hline 
4. & $10 sin(\pi X_1X_2)+20(X_3-0.5)^2+10X_4+5X_5$ & $2.3X_1-3X_2X_4 $  & $X_1,X_2, X_3, X_4,X_5\stackrel{ind}{\sim} U(0,1)$\\
\hline   
\end{tabular} 
\end{adjustbox}
\caption{Table of simulation settings. \label{tab:settings}}
\end{table} 
We set variance standardization ratios of $\alpha = 0.2$, $\delta = 0.45$, and $\gamma = \{0,0.25,0.44\}$, and used the approach described in Section \ref{sim:design} to solve for values of (i) $\beta_1, \beta_2$, $c$ with $\sigma=1$ for a continuous response and (ii)$\beta_0$, $\beta_1, \beta_2$, $c$ for a binary response. Given these solutions and the functions $f,g,h$, we generate the outcome $Y$ or latent values $l(Y)$ as described in the introduction of Section \ref{sims} depending on whether the response is continuous or binary.  Note that $\gamma$ values represent increasing strength of the interaction between the two sets of covariates, and are bounded by $\delta$. Hence, $\gamma$ is zero for additive models and $\gamma$ is non-zero for nonadditive models. The $\gamma=0$ setting results in $\beta_2=0$ and $c=1$ indicating that data is generated from an additive model.

We considered sample sizes of $N = 100, 500$ for continuous outcomes and $N=1000, 3000$ for binary responses. For each scenario, we fit an additive and a nonadditive model to the data, and calculated their respective LPMLs. Then, for $\gamma>0$ we computed the pseudo Bayes factor as $PsBF = exp(LPML_{nonadd}-LPML_{add})$ and for $\gamma=0$ we computed the pseudo Bayes factor as $PsBF = exp(LPML_{add}-LPML_{nonadd})$ so that $PsBF>1$ reflects correct decision. To show the performance of LPML in detecting nonadditivity, we repeated the calculations 200 times for each scenario. In a similar manner, we calculated the OSPEs for the two competing models, and computed their ratio as $R_{OSPE} = \frac{OSPE_{nonadd}}{OSPE_{add}}$ with $R_{OSPE}>1$ reflecting support for additivity.  

\noindent

\subsection{Simulation Results}
In this section, we discuss simulation results. Section \ref{lpml_perf} demonstrates the performance of LPML in detecting additivity in BART models using boxplots. In section \ref{lpml_ospe}, we compare the performance of LPML to that of OSPE with regard to their probability of correctly assessing whether data came from an additive or nonadditive model. 

\subsubsection{Performance of LPML in detecting nonadditivity}
\label{lpml_perf}

\begin{figure}[H]
\begin{center}
 \includegraphics[scale=0.41]{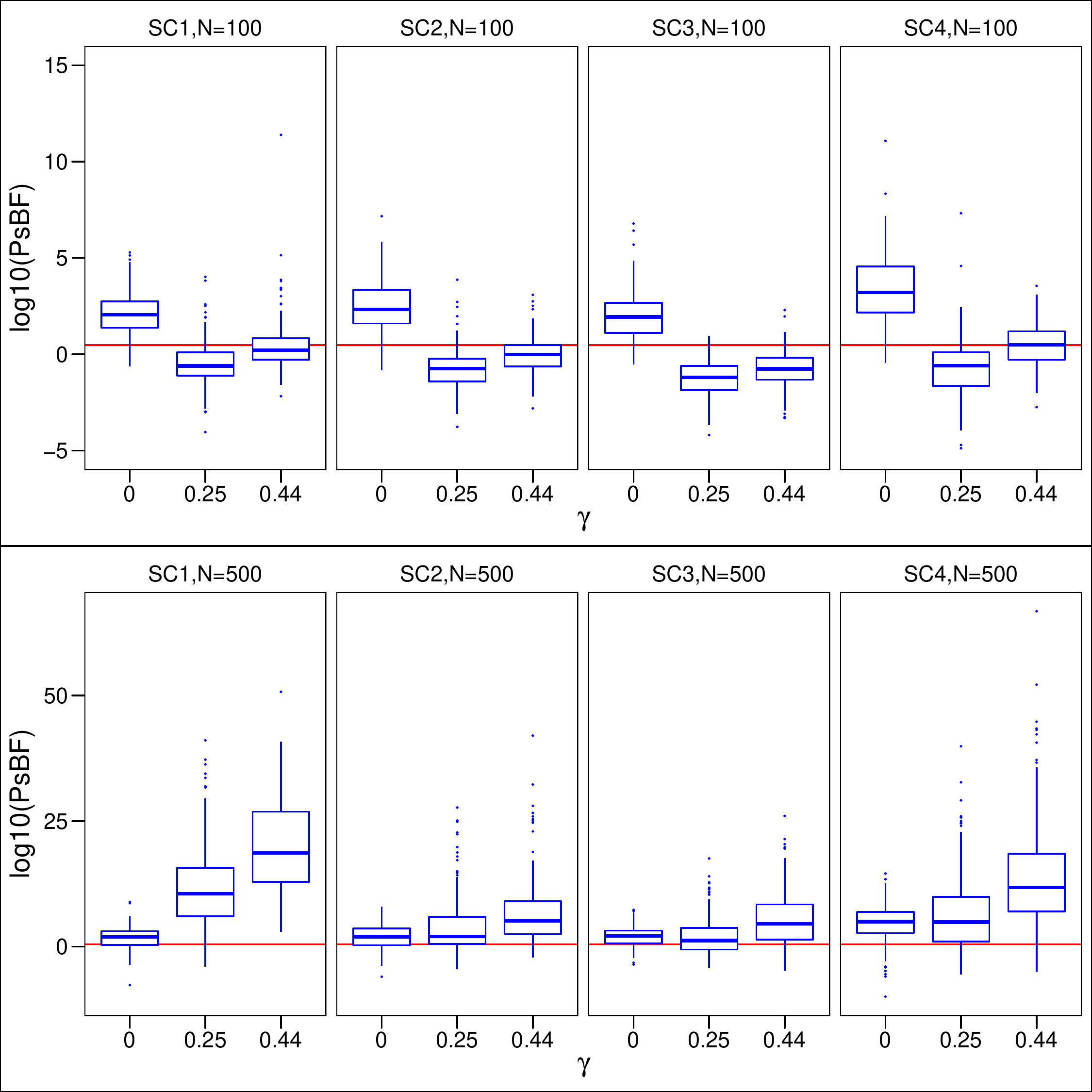}
  \caption{Boxplots of $\log_{10}$(PsBF) for detecting nonadditivity ($\gamma\ne 0$) and additivity ($\gamma=0$) \label{fig:pbfs_2b}} 
\end{center} 
\end{figure}

\begin{figure}[H]
\begin{center}
 \includegraphics[scale=0.41]{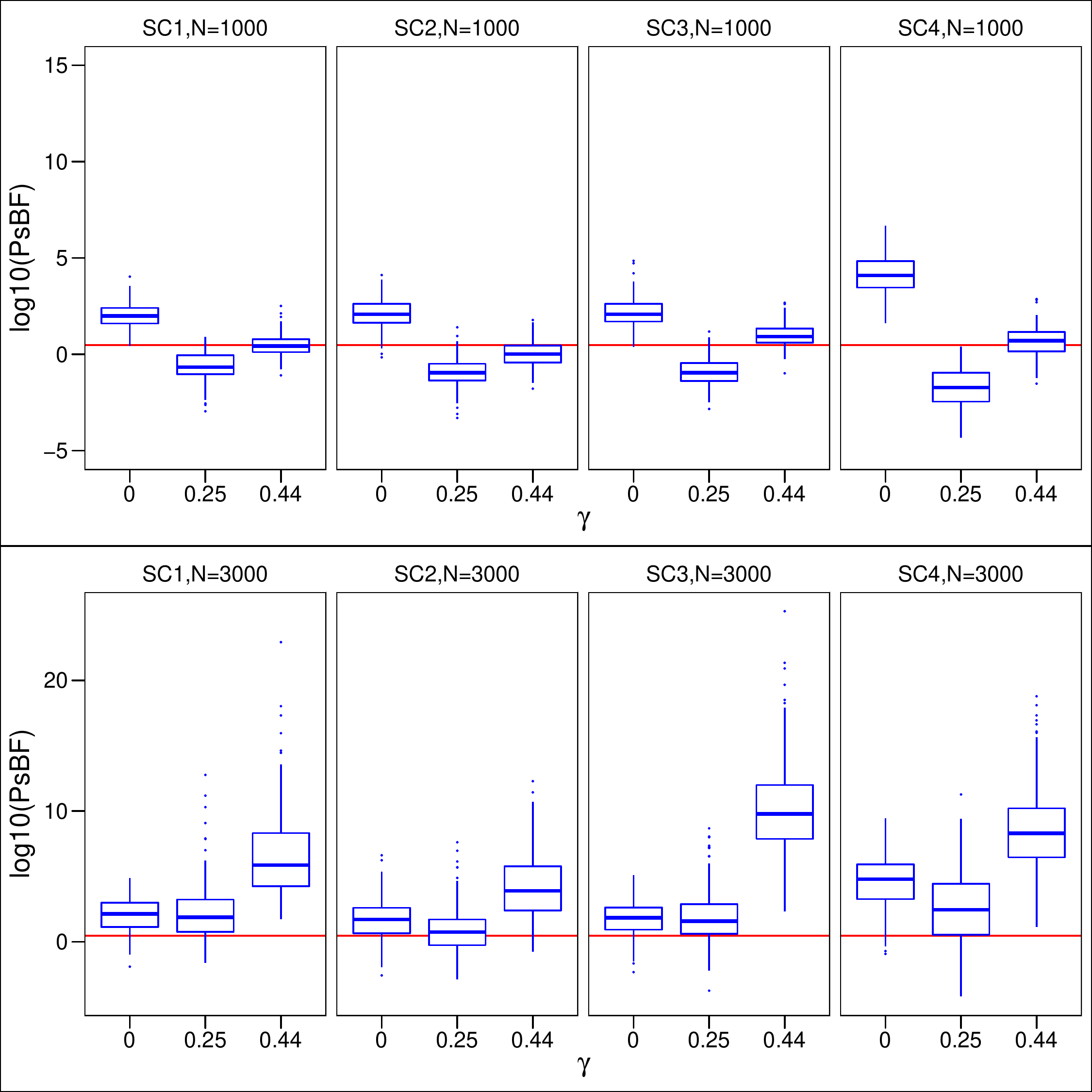}
 \caption{Boxplots of $\log_{10}$(PsBF) for detecting nonadditivity ($\gamma\ne 0$) and additivity ($\gamma=0$) for a binary outcome\label{fig:pbfs_2LB}}
\end{center}
 \end{figure}

\begin{figure}[H]
\begin{center}
 \includegraphics[scale=0.41]{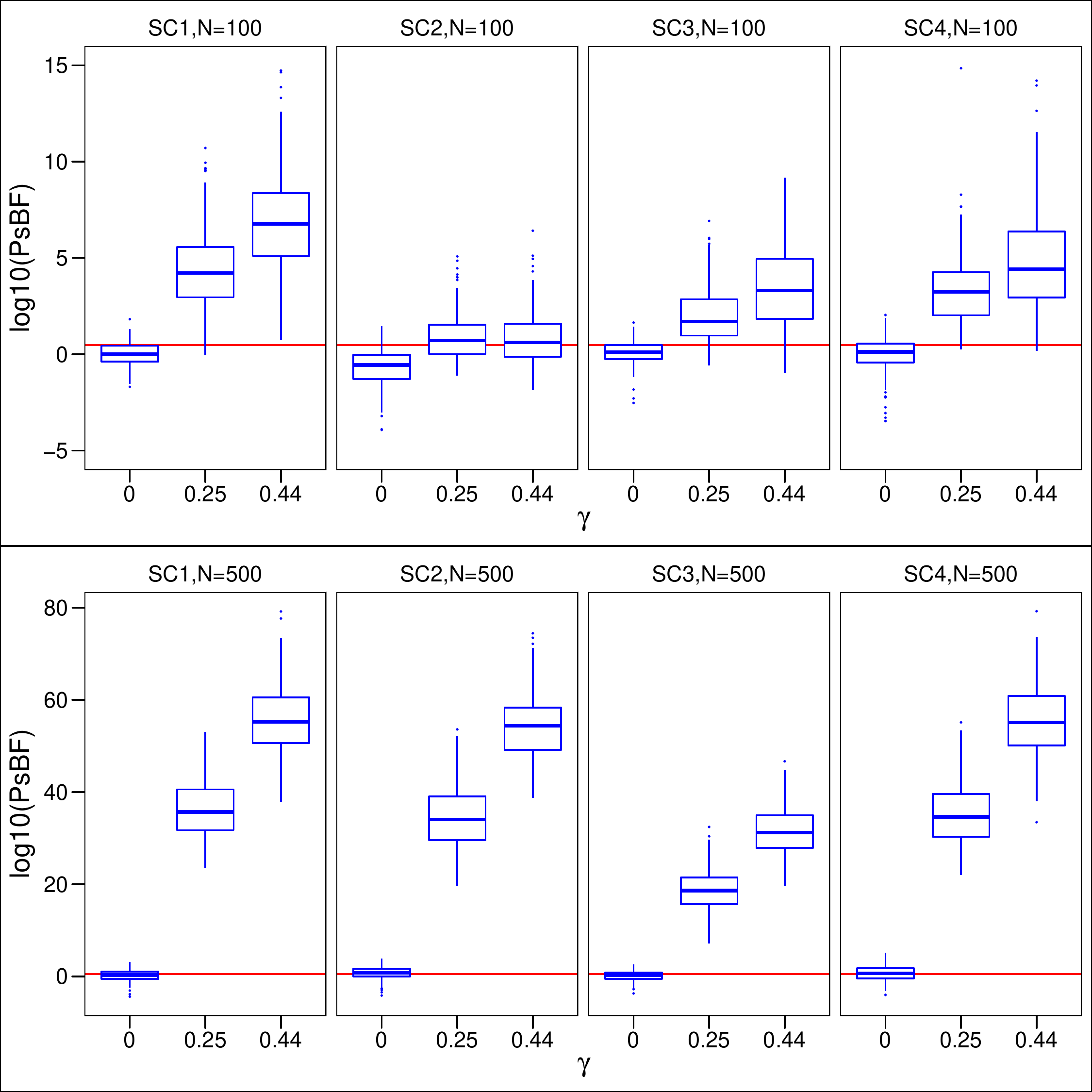}
 \caption{Boxplots of $\log_{10}$(PsBF) for detecting nonadditivity ($\gamma\ne 0$) and additivity ($\gamma=0$) using additive treatment BART model for a continuous outcome \label{fig:pbfs_1bt}}
\end{center}
 \end{figure}

First, we investigate how LPML performs in detecting nonadditivity using BART models. In Figures \ref{fig:pbfs_2b} - \ref{fig:pbfs_1bt} we show boxplots of $\log_{10}(PsBF)$ calculated using 200 datasets in each setting. The ratio of PML's in the PsBF are arranged such that values of $\log_{10}(PsBF)$ greater than $0$ are commensurate with the true model. A red line is drawn at $\log_{10}(3)$ to separate subtantial or stronger evidence from the rest.

In figure \ref{fig:pbfs_2b} for $\gamma=0$, the boxplots of $log_{10}$PsBFs  are above the $\log_{10}(3)$ indicating substantial to decisive evidence in favor of the additive model with most repeated datasets. Notice that generally there is strong evidence in favor of the additive model for $\gamma = 0$ even when the sample size is relatively small. The results are consistent across the various scenarios (SC1 through SC4) investigated.\\

For non-zero $\gamma$ in figure \ref{fig:pbfs_2b}, we are not able to detect nonadditivity when the sample size is small ($N=100$) even in the presence of the strongest setting for interaction with $\gamma=0.44$. This is mainly due to the fact that the nonadditive model is more complex and requires a larger sample size for the function to be accurately estimated. For the larger sample size $N=500$, as $\gamma$ increases, the $\log_{10}$PsBFs also increase providing strong to decisive evidence in support of the nonadditive model as the strength of the interaction term increases. 

In figure \ref{fig:pbfs_2LB}, we show boxplots of $\log_{10}$PsBFs for additive BARTs models with binary response. For $\gamma=0$, the boxplots of $\log_{10}$PsBFs are above the $\log_{10}(3)$ line showing substantial to decisive evidence in favor of the additive model in both small sample size ($N=1000$) and larger sample size ($N=3000$). For non-zero $\gamma$, there is no substantial evidence in favor of nonadditivity with the smaller sample size. For $N=3000$ however, there is increasing evidence in favor of the nonadditive model as the value of $\gamma$ increases.\\
 
In figure \ref{fig:pbfs_1bt}, we present results for the additive treatment BART model with a continuous outcome. For non-zero $\gamma$, there is substantial to decisive evidence in favor of non-additivity for all the scenarios investigated. The strengh of the evidence increases as the sample size and $\gamma$ increase. For $\gamma = 0$, we often see values in the no-decision range between $log_{10}(3)$ and $log_{10}(1/3)$ and very close to zero for the larger sample size. Thus LPML is not able to distinguish between additivity and non-additivity. This may be due to the fact that neither model is wrong, and the single BART nonadditive model, although allowing for interaction, is also able to ignore it to closely approximate the functional relationship of the additive model.

 \subsubsection{Comparing the performance of LPML and OSPE in detecting nonadditivity}
\label{lpml_ospe}

\begin{figure}[H]
\begin{center}
  \includegraphics[scale=0.7]{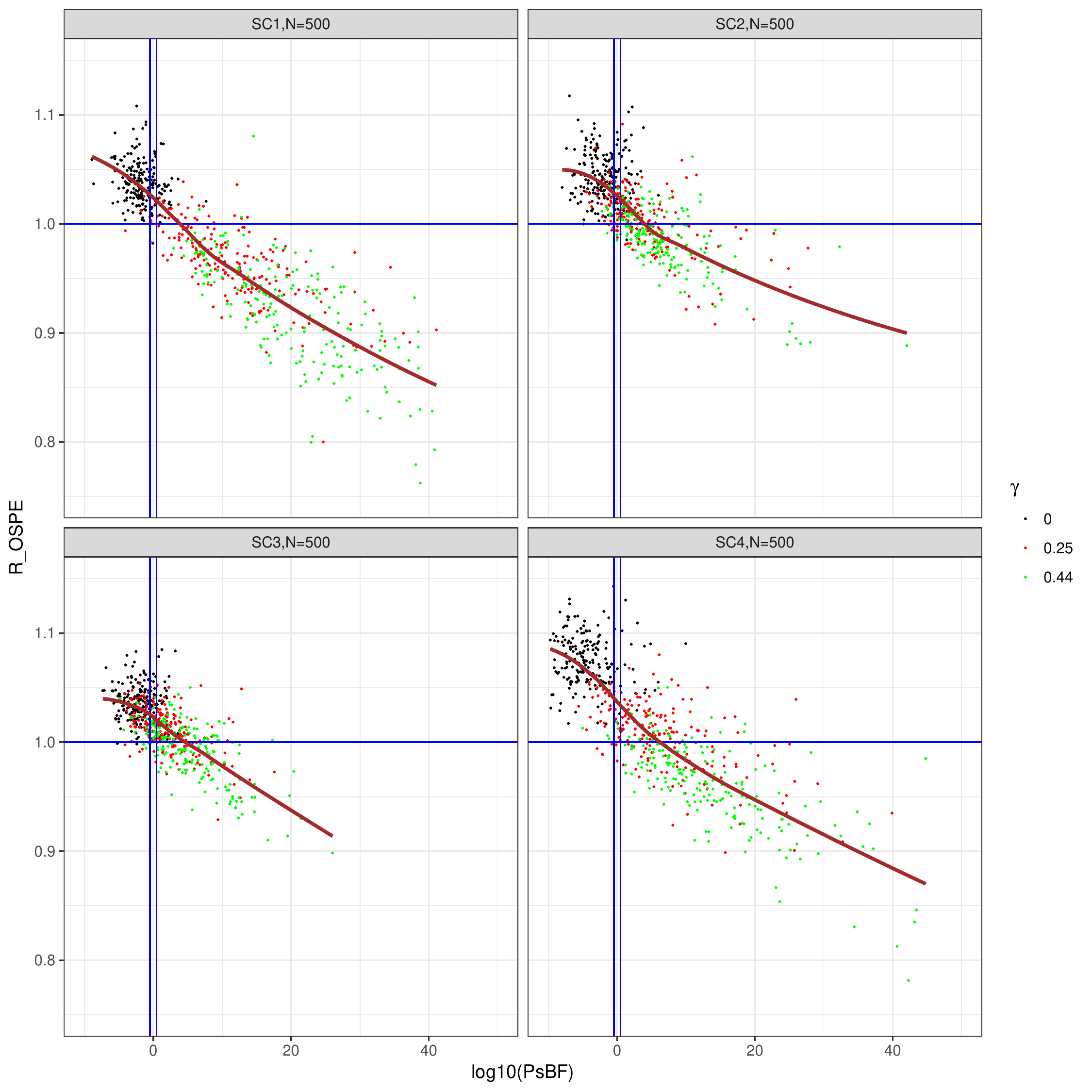}
\caption{Scatterplots of $\log_{10}(PsBF)$  against a ratio of OSPE, $N=500$, for four scenarios, dot colors indicating values of $\gamma$.\label{fig:scatt_pbfospe}}
\end{center} 
\end{figure}

Next we compare the performance of LPML and OSPE criteria in assessing nonadditivity. In Figure \ref{fig:scatt_pbfospe}, we plot a scatter plot of ratios of OSPEs against the $\log_{10}$PsBFs for each dataset for $N=500$ to show how the two criteria compare in assessing nonadditivity. In this plot, we compute these criteria, pitting nonadditive models versus additive, so that $PsBF = exp(LPML_{nonadd}-LPML_{add})$ regardless of $\gamma$, in contrast to the prior definition.  A model with lower OSPE is preferred and decisions for LPML are based on Jeffreys' recommendations. For additive models with $\gamma=0$ (blue dots), Figure \ref{fig:scatt_pbfospe} shows OSPE able to correctly classify almost all additive models as seen by the ratios of OSPEs being above 1.  In comparison, LPML is also able to correctly classify most of the additive models as shown by the points on the left of the $\log_{10}(1/3)$ line, with a few points incorrectly classified as nonadditive. For nonadditive models, LPML correctly identifies most of them as nonadditive as shown by points on the right of the $\log_{10}(3)$ line. OSPE is also able to detect nonadditivity well, but has more misclassified points compared to LPML. The results are consistent accross the different scenarios and other sample sizes (not shown) examined. As mentioned before, LPML provides an interpretable quantification of evidence in favor of nonadditivity versus additivity.

\begin{figure}[H] 
 \includegraphics[scale=0.7]{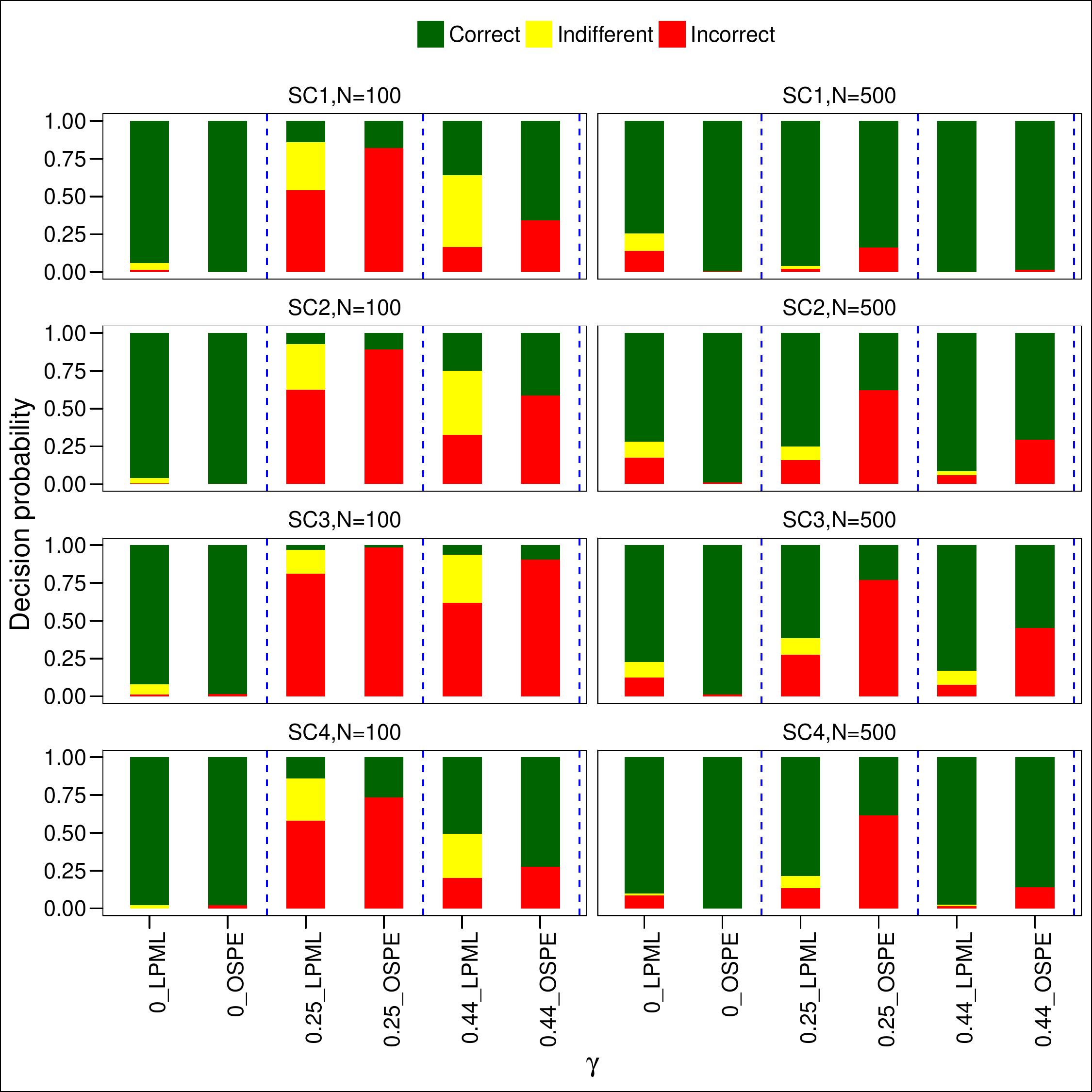}
\caption{Probabilities of decision outomes with two sets of covariates, continuous response \label{fig:probs_2b}} 
\end{figure}

\begin{figure}[H]
\begin{center}
 \includegraphics[scale=0.7]{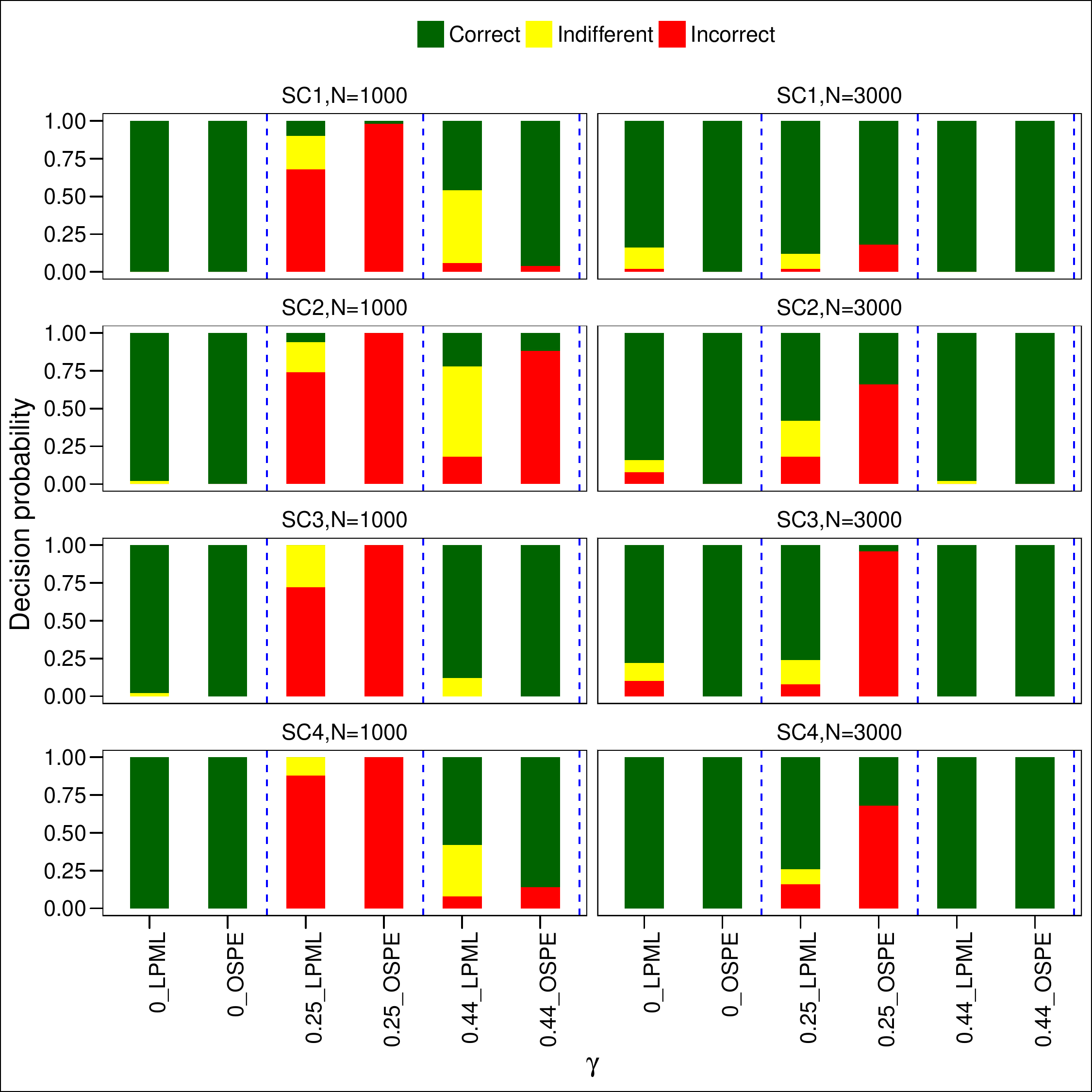}
 \caption{Probability of decision outomes with two sets of covariates, binary response\label{fig:probs_2LB}}
\end{center}
 \end{figure}

\begin{figure}[H]
\begin{center}
 \includegraphics[scale=0.7]{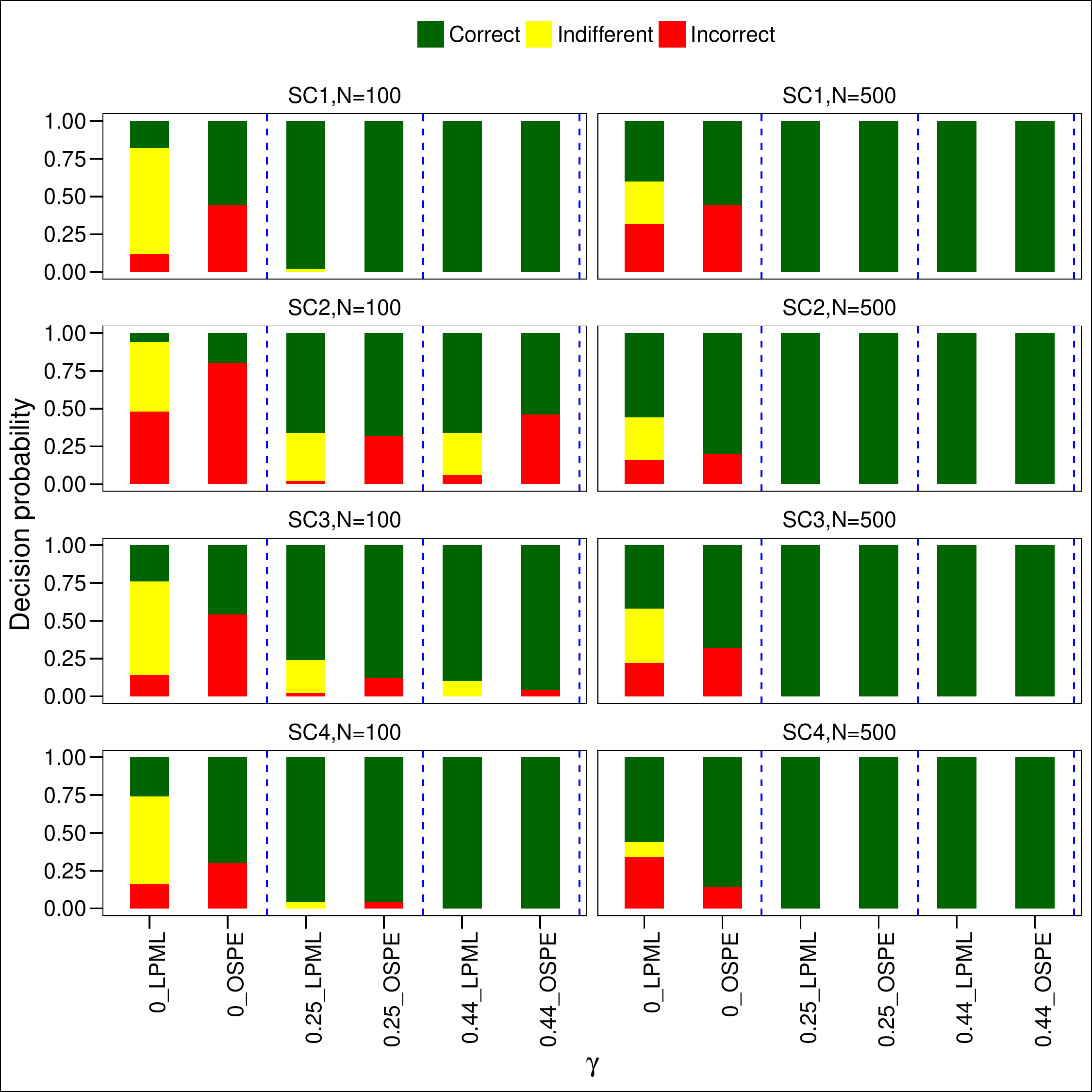}
 \caption{Probability of decision outomes with a treatment and covariates,continuous response \label{fig:probs_1bt}}
\end{center}
 \end{figure}

Figure \ref{fig:probs_2b} depicts the probability of choosing the correct model using LPML versus OSPE for continuous response. We use a cutoff value of 1 for the ratio of  OSPEs, and threshold values given by Jeffreys \citep{jeffreys1961} for LPML. Thus, we classify a decision as indifferent if the PsBF is between $1/3$ and $3$. For the smaller sample size of $N=100$, both criteria identify the additive model correctly, but fail to identify the nonadditive model with OSPE having the higher rates of misclassified models. As the sample size increases to $N=500$, OSPE correctly identifies additive models with higher probability of correctly identified models compared to that of LPML. However, LPML outperforms OSPE in identifying the nonadditive models with higher rate of correctly identified nonadditive models compared to that of OSPE. For both criteria, the probability of correctly identifying the nonadditive model increases as $\gamma$ and the sample size increase, accross all the scenarios. We observe similar patterns in Figure \ref{fig:probs_2LB} when the outcome is binary. For the additive treatment case with continuous response in Figure~\ref{fig:probs_1bt}, OSPE outperforms LPML in idenfifying additive models, with LPML making an indifferent decision in many cases. However, LPML outperforms OSPE in correctly identifying the nonadditive models. The probability of correctly identifying the models increase as the sample size increases. 
  
\section{Application of additive BARTs model to Type 2 diabetes data and transplant data}
\label{sec:application}
\subsection{Type 2 diabetes data}
\label{sec:type2}
Consider the Type 2 diabetes example introduced in Section \ref{sec:intro}. Fitting a single BART model with health and demographic variables makes the model very complex and requires considerable post processing to extract interpretable, individual covariates' effects. If the health and demographic effects are additive, the new additive model is simpler and requires less post processing. We fit additive and nonadditive models to health and demographic variables  with A1C at 6 months as the outcome.
 \[\text{Nonadditive: } A1C6 = f(health,demog)+\epsilon 
 \] 
 \[
 \text{ Additive: } A1C6 = f(health) + f(demog)+ \epsilon  \]
 The resulting PsBF comparing the nonadditive vesus additive model is $25638.04$ indicating decisive  evidence in favor of the nonadditive model. We next test whether the treatment effect is additive. We fit the following models
 \[ 
 \text{Nonadditive: } A1C6 = f(Trt,health,demog)+\epsilon 
 \] 
\[
 \text{ Additive: } A1C6 = \beta Trt + f(health,demog)+ \epsilon  
 \] 
 We obtain PsBF of $\num{0.09521}$ when comparing the nonadditive versus additive model, indicating strong evidence in favor of the additive model. Posterior mean for the treatment effect is $-0.0769$ with a $95\%$ credible interval of ($-0.1903,0.0344$). The overall treatment effect is directly estimated and easy to interpret as a constant effect across the population. 
 
\subsection{Hematopoietic Cell Transplant (HCT) data}
\label{sec:transplant}

We also illustrate the proposed methodology with a study reported in \citep{LogaSpar17}, of one year survival outcomes after a hematopoietic cell transplant (HCT) that is used to treat a variety of hematologic malignancies. We use data of 3802 patients receiving reduced intensity conditioning for their HCT between 2011 and 2013, with data reported to the Center for International Blood and Marrow Transplant Research (CIBMTR) registry. Follow up of one year is complete for all patients, so we analyze the outcomes using binary methods with outcome defined as the patient's survival status at one year. The primary treatment of interest is the type of conditioning regimen used (Fludarabine/Melphalan, or FluMel for short, vs. Fludarabine/Busulfan, or FluBu for short). A variety of patient, donor, and disease factors were examined for their utility in personalizing the selection of the conditioning regimen, including age, race/ethnicity, performance score, Cytomegalovirus status, disease, remission status, disease subtypes, chemosensitivity, interval from diagnosis to transplant, donor type, Human Leukocyte Antigen (HLA) matching between donor and recipient, prior autologous transplant, gender matching between donor and recipient, comorbidity score, and year of transplant. This observational cohort appears to be well balanced between the regimens across these factors, indicating reasonable equipoise by clinicians on which conditioning regimen is most appropriate for individual patients. Logan \textit{et al.} in \citep{LogaSpar17} used BART to estimate the treatment difference in one year survival and used a tree approximation to the BART fit for purposes of interpretation. Our goal is to assess whether the effect of the treatment on one year survival is additive with respect to the rest of the factors. We compared the following models

 \[ 
 \text{Nonadditive: } l(Y) = f(Trt,patient, donor,disease factors) 
 \] 
\[
 \text{ Additive: } l(Y)  = \beta Trt + f(patient, donor,disease factors) \ .
 \] 

Taking the approach in sections \ref{sec:bart_bin} and \ref{onebt:binary}, the resulting PsBF for comparing nonadditive versus additive model is $2173.34$ indicating decisive evidence in favor of the nonadditive model. This implies that treatment assignment depends on each patient's characteristics, the same conclusion reached in \citep{LogaSpar17}.

\section{Conclusion and discussion}
\label{sec:conc}
 
In this article, we proposed an approach for assessing additivity in Bayesian additive regression trees using a ratio of pseudo marginal likelihoods. We used BART as our building block for the additive models since it provides a flexible, fully nonparametric way to approximate complex functional relationships including nonlinearities and interactions. The proposed additive BARTs models with additive components provided good predictive models while allowing for simpler BARTs models to be used if appropriate. In the case of the additive treatment BART model, the resulting treatment effect estimate is readily available from the model fit and interpretable directly.
 
To assess whether a model is additive or not, we propose using the pseudo Bayes factor (ratio of PML's) and out-of-sample predictive error criteria.  The former provided two main advantages over the latter; it is very fast to compute using posterior samples and has related threshold values to measure the magnitude of the evidence. The OSPE requires computationally intensive cross-validation. Both criteria provided good performance in identifying the correct models. However, LPML outperformed OSPE in identifying nonadditive models in many instances as shown in our simulation results.
 
 We observed that the evidence provided by the LPML criterion in favor of nonadditivity increased considerably with increasing sample size and increasing strength of the interaction term ($\gamma$) for all cases considered. LPML was also able to detect additivity when a model with two additive BARTs components was used. However, additivity of treatment effect was not readily detectable when the additive treatment model was used. One explanation for this is that a single BART is able to accurately recover the additive functional relationship, resulting in both single BART and additive treatment BART model being accurate. Further investigation and post processing are recommended to assess closeness of estimates from these two models. 
 
%

\begin{thebibliography}{40}
\providecommand{\natexlab}[1]{#1}
\providecommand{\url}[1]{\texttt{#1}}
\expandafter\ifx\csname urlstyle\endcsname\relax
  \providecommand{\doi}[1]{doi: #1}\else
  \providecommand{\doi}{doi: \begingroup \urlstyle{rm}\Url}\fi

\bibitem[Abbot and Marohasy(2017)]{AbboMaro17}
John Abbot and Jennifer Marohasy.
\newblock Application of artificial neural networks to forecasting monthly
  rainfall one year in advance for locations within the murray darling basin,
  australia.
\newblock \emph{International Journal of Sustainable Development and Planning},
  12\penalty0 (8):\penalty0 1282--1298, 2017.
\newblock ISSN 1743-761X.

\bibitem[Albert and Chib(1993)]{AlbeChib93}
James~H. Albert and Siddhartha Chib.
\newblock Bayesian analysis of binary and polychotomous response data.
\newblock \emph{Journal of the American Statistical Association}, 88\penalty0
  (422):\penalty0 669--679, 1993.
\newblock ISSN 01621459.
\newblock URL \url{http://www.jstor.org/stable/2290350}.

\bibitem[Ansari and Jedidi(2000)]{AnsaJedi00}
Asim Ansari and Kamel Jedidi.
\newblock Bayesian factor analysis for multilevel binary observations.
\newblock \emph{Psychometrika}, 65\penalty0 (4):\penalty0 475--496, 2000.
\newblock ISSN 0033-3123.

\bibitem[Association(2014)]{Ada14}
American~Diabetes Association.
\newblock Standards of medical care in diabetes.
\newblock \emph{Diabetes Care}, 37\penalty0 (3):\penalty0 887--887, 2014.
\newblock ISSN 0149-5992.

\bibitem[Barcella et~al.(2017)Barcella, De~Iorio, Favaro, and
  Rosner]{BarcDeIo17}
William Barcella, Maria De~Iorio, Stefano Favaro, and Gary~L Rosner.
\newblock Dependent generalized dirichlet process priors for the analysis of
  acute lymphoblastic leukemia.
\newblock \emph{Biostatistics}, 2017.

\bibitem[Barry(1993)]{barry1993}
Daniel Barry.
\newblock Testing for additivity of a regression function.
\newblock \emph{The Annals of Statistics}, 21\penalty0 (1):\penalty0 235--254,
  1993.

\bibitem[Bhattacharya and Sengupta(2009)]{BhatSeng09}
Sourabh Bhattacharya and Ashis Sengupta.
\newblock Bayesian analysis of semiparametric linear-circular models.
\newblock \emph{Journal of agricultural, biological, and environmental
  statistics}, 14\penalty0 (1):\penalty0 33, 2009.
\newblock ISSN 1085-7117.

\bibitem[Booth et~al.(2011)Booth, Eilertson, Olinares, and Yu]{BootEile11}
James~G Booth, Kirsten~E Eilertson, Paul Dominic~B Olinares, and Haiyuan Yu.
\newblock A bayesian mixture model for comparative spectral count data in
  shotgun proteomics.
\newblock \emph{Molecular \& Cellular Proteomics}, 10\penalty0 (8):\penalty0
  M110. 007203, 2011.
\newblock ISSN 1535-9476.

\bibitem[Branscum et~al.(2015)Branscum, Johnson, Hanson, and Baron]{BranJohn15}
Adam~J Branscum, Wesley~O Johnson, Timothy~E Hanson, and Andre~T Baron.
\newblock Flexible regression models for roc and risk analysis, with or without
  a gold standard.
\newblock \emph{Statistics in medicine}, 34\penalty0 (30):\penalty0 3997--4015,
  2015.
\newblock ISSN 1097-0258.

\bibitem[Cardoso and Tempelman(2004)]{CardTemp04}
FF~Cardoso and RJ~Tempelman.
\newblock Hierarchical bayes multiple-breed inference with an application to
  genetic evaluation of a nelore-hereford population 1.
\newblock \emph{Journal of Animal Science}, 82\penalty0 (6):\penalty0
  1589--1601, 2004.
\newblock ISSN 1525-3163.

\bibitem[Chang et~al.(2006)Chang, Gianola, Heringstad, and
  Klemetsdal]{ChanGian06}
Y~M Chang, D~Gianola, B~Heringstad, and G~Klemetsdal.
\newblock A comparison between multivariate slash, student's t and probit
  threshold models for analysis of clinical mastitis in first lactation cows.
\newblock \emph{Journal of Animal Breeding and Genetics}, 123\penalty0
  (5):\penalty0 290--300, 2006.
\newblock ISSN 1439-0388.

\bibitem[Chang et~al.(2004)Chang, Gianola, Heringstad, and
  Klemetsdal]{ChanGian04}
Yu-Mei Chang, Daniel Gianola, Bjørg Heringstad, and Gunnar Klemetsdal.
\newblock Longitudinal analysis of clinical mastitis at different stages of
  lactation in norwegian cattle.
\newblock \emph{Livestock Production Science}, 88\penalty0 (3):\penalty0
  251--261, 2004.
\newblock ISSN 0301-6226.

\bibitem[Chen et~al.(2002)Chen, Ibrahim, and Lipsitz]{ChenIbra02}
Ming-Hui Chen, Joseph~G Ibrahim, and Stuart~R Lipsitz.
\newblock Bayesian methods for missing covariates in cure rate models.
\newblock \emph{Lifetime Data Analysis}, 8\penalty0 (2):\penalty0 117--146,
  2002.
\newblock ISSN 1380-7870.

\bibitem[Chen et~al.(2014)Chen, Hanson, and Zhang]{ChenHans14}
Yuhui Chen, Timothy Hanson, and Jiajia Zhang.
\newblock Accelerated hazards model based on parametric families generalized
  with bernstein polynomials.
\newblock \emph{Biometrics}, 70\penalty0 (1):\penalty0 192--201, 2014.
\newblock ISSN 1541-0420.

\bibitem[Cheng et~al.(2017)Cheng, Gill, Choi, Zhou, Jia, and Xie]{ChenGill17}
Wen Cheng, Gurdiljot~Singh Gill, Simon Choi, Jiao Zhou, Xudong Jia, and Meiquan
  Xie.
\newblock Comparative evaluation of temporal correlation treatment in crash
  frequency modelling.
\newblock \emph{Transportmetrica A: Transport Science}, \penalty0
  (just-accepted):\penalty0 1--40, 2017.
\newblock ISSN 2324-9935.

\bibitem[Chipman and McCulloch(2016)]{ChipMacC16}
Hugh Chipman and Robert McCulloch.
\newblock \emph{BayesTree: Bayesian Additive Regression Trees}, 2016.
\newblock R package version 0.3-1.4.

\bibitem[Chipman et~al.(2010)Chipman, George, and {E. McCulloch}]{bart2010}
Hugh~A. Chipman, Edward~I. George, and Robert {E. McCulloch}.
\newblock Bart: Bayesian additive regression trees.
\newblock \emph{The Annals of Applied Statistics}, 4\penalty0 (1):\penalty0
  266--298, 2010.
\newblock \doi{10.1214/09-AOAS285}.

\bibitem[Christensen et~al.(2011)Christensen, Johnson, Branscum, and
  Hanson]{ChrisJohn11}
Ronald Christensen, Wesley Johnson, Adam Branscum, and Timothy~E. Hanson.
\newblock \emph{Bayesian ideas and data analysis: an introduction for
  scientists and statisticians}.
\newblock CRC Press, 2011.

\bibitem[Dey et~al.(1997)Dey, Chen, and Chang]{DeyChen97}
Dipak~K Dey, Ming-Hui Chen, and Hong Chang.
\newblock Bayesian approach for nonlinear random effects models.
\newblock \emph{Biometrics}, pages 1239--1252, 1997.
\newblock ISSN 0006-341X.

\bibitem[Eubank et~al.(1995)Eubank, Hart, Simpson, and Stefanski]{eubank1995}
R.~L. Eubank, J.~D. Hart, D.~G. Simpson, and L.~A. Stefanski.
\newblock Testing for additivity in nonparametric regression.
\newblock \emph{The Annals of Statistics}, 23\penalty0 (6):\penalty0
  1896--1920, 1995.

\bibitem[Geisser and Eddy(1979)]{GeisEddy79}
Seymour Geisser and William~F. Eddy.
\newblock A predictive approach to model selection.
\newblock \emph{Journal of the American Statistical Association}, 74\penalty0
  (365):\penalty0 153--160, 1979.

\bibitem[Gelfand and Dey(1994)]{GelfandDey1994}
Alan~E. Gelfand and D.~K. Dey.
\newblock Bayesian model choice: Asymptotics and exact calculations.
\newblock \emph{Journal of the Royal Statistical Society. Series B
  (Methodological)}, 56\penalty0 (3):\penalty0 501--514, 1994.

\bibitem[Gilks et~al.(1995)Gilks, Richardson, and Spiegelhalter]{Gelf1995}
Walter~R Gilks, Sylvia Richardson, and David Spiegelhalter.
\newblock \emph{Markov chain Monte Carlo in practice}.
\newblock CRC press, 1995.

\bibitem[Han and Carlin(2001)]{HanCarl2001}
Cong Han and Bradley~P Carlin.
\newblock Markov chain monte carlo methods for computing bayes factors.
\newblock \emph{Journal of the American Statistical Association}, 96\penalty0
  (455):\penalty0 1122--1132, 2001.
\newblock \doi{10.1198/016214501753208780}.
\newblock URL \url{https://doi.org/10.1198/016214501753208780}.

\bibitem[Hanson and Yang(2007)]{HansTim2007}
Timothy Hanson and Mingan Yang.
\newblock Bayesian semiparametric proportional odds models.
\newblock \emph{Biometrics}, 63\penalty0 (1):\penalty0 88--95, 2007.
\newblock ISSN 1541-0420.

\bibitem[Hanson et~al.(2011)Hanson, Branscum, and Johnson]{HansTim2011}
Timothy~E Hanson, Adam~J Branscum, and Wesley~O Johnson.
\newblock Predictive comparison of joint longitudinal-survival modeling: a case
  study illustrating competing approaches.
\newblock \emph{Lifetime data analysis}, 17\penalty0 (1):\penalty0 3--28, 2011.
\newblock ISSN 1380-7870.

\bibitem[Heydari et~al.(2016)Heydari, Fu, Lord, and Mallick]{HeydFu16}
Shahram Heydari, Liping Fu, Dominique Lord, and Bani~K Mallick.
\newblock Multilevel dirichlet process mixture analysis of railway grade
  crossing crash data.
\newblock \emph{Analytic methods in accident research}, 9:\penalty0 27--43,
  2016.
\newblock ISSN 2213-6657.

\bibitem[Holmes et~al.(2006)Holmes, Held, et~al.]{HolmHeld06}
Chris~C Holmes, Leonhard Held, et~al.
\newblock Bayesian auxiliary variable models for binary and multinomial
  regression.
\newblock \emph{Bayesian analysis}, 1\penalty0 (1):\penalty0 145--168, 2006.

\bibitem[Hu et~al.(2015)Hu, Huang, and Tiwari]{HuHuang2015}
Na~Hu, Lan Huang, and Ram~C Tiwari.
\newblock Signal detection in fda aers database using dirichlet process.
\newblock \emph{Statistics in medicine}, 34\penalty0 (19):\penalty0 2725--2742,
  2015.
\newblock ISSN 1097-0258.

\bibitem[Jeffreys(1961)]{jeffreys1961}
H~Jeffreys.
\newblock Theory of probability,(oxford: Oxford university press).
\newblock 1961.

\bibitem[Kapelner and Bleich(2016)]{KapelBleich16}
Adam Kapelner and Justin Bleich.
\newblock {bartMachine}: Machine learning with {B}ayesian additive regression
  trees.
\newblock \emph{Journal of Statistical Software}, 70\penalty0 (4):\penalty0
  1--40, 2016.

\bibitem[Kass and Raftery(1995)]{KassRaft1995}
Robert~E. Kass and Adrian~E. Raftery.
\newblock Bayes factors.
\newblock \emph{Journal of the American Statistical Association}, 90\penalty0
  (430):\penalty0 773--795, 1995.
\newblock \doi{10.1080/01621459.1995.10476572}.
\newblock URL
  \url{https://amstat.tandfonline.com/doi/abs/10.1080/01621459.1995.10476572}.

\bibitem[Koenig(1980)]{KoenCera80}
R~J Koenig.
\newblock Hemoglobin aic and diabetes mellitus.
\newblock \emph{Annual Review of Medicine}, 31\penalty0 (1):\penalty0 29--34,
  1980.

\bibitem[Li et~al.(2006)Li, Bolt, and Fu]{LiBolt06}
Yanmei Li, Daniel~M Bolt, and Jianbin Fu.
\newblock A comparison of alternative models for testlets.
\newblock \emph{Applied Psychological Measurement}, 30\penalty0 (1):\penalty0
  3--21, 2006.
\newblock ISSN 0146-6216.

\bibitem[Linero(2018)]{Line18}
Antonio~R Linero.
\newblock Bayesian regression trees for high-dimensional prediction and
  variable selection.
\newblock \emph{Journal of the American Statistical Association}, pages 1--11,
  2018.

\bibitem[Logan et~al.(2017)Logan, Sparapani, Mcculloch, and Laud]{LogaSpar17}
Brent~R Logan, Rodney Sparapani, Robert~E Mcculloch, and Purushottam~W Laud.
\newblock Decision making and uncertainty quantification for individualized
  treatments using bayesian additive regression trees.
\newblock \emph{Statistical Methods in Medical Research}, pages 1--15, 2017.
\newblock \doi{10.1177/0962280217746191}.

\bibitem[McCulloch et~al.(2017)McCulloch, Sparapani, Gramacy, Pratola, and
  Spanbauer]{bartmanual}
Robert McCulloch, Rodney Sparapani, Robert Gramacy, Matthew Pratola, and
  Charles Spanbauer.
\newblock \emph{BART: Bayesian Additive Regression Trees}, 2017.
\newblock R package version 1.4.

\bibitem[O{'}Malley et~al.(2003)O{'}Malley, Normand, and Kuntz]{OMalNorm03}
A~James O{'}Malley, Sharon{E}lise~T Normand, and Richard~E Kuntz.
\newblock Application of models for multivariate mixed outcomes to medical
  device trials: coronary artery stenting.
\newblock \emph{Statistics in medicine}, 22\penalty0 (2):\penalty0 313--336,
  2003.
\newblock ISSN 1097-0258.

\bibitem[Sparapani et~al.(2016)Sparapani, Logan, McCulloch, and
  Laud]{SparLoga16}
Rodney~A Sparapani, Brent~R Logan, Robert~E McCulloch, and Purushottam~W Laud.
\newblock Nonparametric survival analysis using bayesian additive regression
  trees (bart).
\newblock \emph{Statistics in medicine}, 35\penalty0 (16):\penalty0 2741--2753,
  2016.

\bibitem[Zhang et~al.(2016)Zhang, Staicu, and Maity]{zhang2016}
Yichi Zhang, Ana-Maria Staicu, and Arnab Maity.
\newblock Testing for additivity in non-parametric regression.
\newblock \emph{Canadian Journal of Statistics}, 44\penalty0 (4):\penalty0
  445--462, Aug 2016.

\end{thebibliography}

\end{document}